\documentclass[fleqn,twoside]{article}
\usepackage{espcrc2}
\usepackage{graphicx}

\newlength{\figwidth}
\renewcommand{\vec}[1]{\mbox{\boldmath $#1$}}

\title{Temporal meson correlators at finite temperature
       on quenched anisotropic lattice%
  \thanks{Poster presented by K. Nomura}}

\author{
 K. Nomura%
  \address{Department of Physics, Hiroshima University, 
           Higashi-Hiroshima 739-8626, Japan \vspace{-0.25cm}},
 O. Miyamura$^{\rm a}$,
 T. Umeda%
  \address{Center for Computational Physics, University 
           of Tsukuba, Tsukuba 305-8577, Japan \vspace{-0.25cm}},
  and
 H. Matsufuru%
  \address{Yukawa Institute for Theoretical
           Physics, Kyoto University, Kyoto 606-8502, Japan} }

\begin{document}

\begin{abstract}
We study charmonium correlators at finite temperature
in quenched anisotropic lattice QCD.
The smearing technique is applied to enhance the 
low energy part of the correlator.
We use two analysis procedures:
the maximum entropy method for extraction of the spectral function
without assuming specific form, as an estimate of the shape of
spectral function, and the $\chi^2$ fit assuming typical
forms as quantitative evaluation of the parameters associated
to the forms.
We find that at $T\simeq 0.9T_c$ the ground state peak has almost
the same mass as at $T\!=\!0$ and almost vanishing width.
At $T\simeq 1.1T_c$, our result suggests that the correlator
still has nontrivial peak structure at almost the same position as
below $T_c$ with finite width.

\vspace{1pc}
\end{abstract}

\maketitle

\section{Introduction}

It is expected that the hadronic excitation modes strongly change
their properties around the QCD phase transition.
Such changes of hadron properties may signal
the occurrence of phase transition in heavy ion collision experiments.
With the potential model approach, it has been expected that
the charmonium masses are shifted in the vicinity of $T_c$
\cite{MASSshift}.
In the plasma phase, they are expected to dissolve,
and resultant suppression of $J/\psi$ formation
has been regarded as one of most important signals of plasma
formation \cite{Jpsi}.
Therefore it is important to verify these phenomena using
lattice QCD in a model independent way.

In principle, information of excitation modes can be extracted 
from the Matsubara Green function in the Euclidean time direction.
In practice, however, there are several problems at finite temperature.
One problem is that the degrees of freedom is restricted
by the short temporal extent.
To avoid this difficulty, we use an anisotropic lattice, on which 
temporal lattice spacing is finer than the spatial one.
The other problem is that one need to extract the low energy structure
from the correlator at the region of $O(1/T)$,
where the correlator contains contribution from wide range
of frequency of the spectral function.
We circumvent this problem by applying the smearing technique,
which enhances the low energy part of the correlator.

In order to extract a reliable information on the spectral function,
we suggest to use maximum entropy method (MEM) \cite{NAH99}
and $\chi^2$ fit method in a complementary manner.
The former has an advantage that it does not require to assume
specific form for the spectral function.
Once the form of spectral function is estimated, however, the latter
approach is more quantitative for evaluation of properties of the mode,
such as mass and width.
We first discuss how the applicability of these procedures to finite
temperatures is justified using correlators at $T\!=\!0$ by varying
numbers of degrees of freedom used in analyses.
Then we discuss changes of spectral function below and above $T_c$.
Details of this work will be presented in future publication.

\section{Lattice setup}

We use quenched lattices of the sizes $20^3\times N_t$, where
$N_t=160$, 32, and 26 which roughly correspond to $T\simeq 0$,
0.9$T_c$,
and 1.1$T_c$, respectively.
The zero temperature lattice, and the setup of lattice parameters
are the same as those used in Ref.~\cite{Aniso01b}.
The gauge configurations are generated with the Wilson plaquette
action at $\beta=6.10$ with the anisotropy $\xi=a_{\sigma}/a_t = 4$.
The spatial cutoff set by the hadronic radius $r_0$ is
$a^{-1}_{\sigma}=2.030(13)$ GeV.
$N_t=28$ is close to the phase transition.
The quark action is the $O(a)$ improved Wilson action \cite{Aniso01b},
with the hopping parameter corresponding to the quark mass
$m_Q=0.98$ GeV.

The correlator is represented as
\vspace{-0.07cm}
\begin{equation}
 C(t)= \sum_{\mbox{\scriptsize\boldmath $x$}}
       \langle O(\vec{x},t) O^{\dag}(0,0) \rangle.
\end{equation}
\vspace{-0.1cm}
The operator $O(\vec{x},t)$ is
\vspace{-0.07cm}
\begin{equation}
O(\vec{x},t)= \sum_{\mbox{\scriptsize\boldmath $y$}}
 \phi(\vec{y}) \bar{q} ( \vec{x} + \vec{y},t) \Gamma q(\vec{x},t),
\end{equation}
\vspace{-0.1cm}
where $4\times 4$ matrix $\Gamma$ specifies the quantum number and 
$\phi(\vec{y})$ is the smearing function for which
we use the wave function observed at $T\!=\!0$.
The correlators are measured on 500 configurations at $T\!=\!0$ and
1000 configurations at $T>0$.
In order to reduce the statistical error, we average 16 correlators
measured with different source points on each configuration.

\section{Analysis procedure}

We focus on the low energy structure of the spectral function.
We use two types of analysis method:
the maximum entropy method (MEM), and the standard $\chi^2$ fit
with ansatz for the shape of spectral function $A(\omega)$.
MEM has an advantage that it does not require to assume a specific
form for the spectral function.
When the number of temporal points is sufficiently large,
and the data region is also in sufficiently long distance,
MEM can reproduce the spectral function successfully
\cite{NAH99,MEMzero}.
However, when the number of temporal points is small, the extracted
spectral function has a large uncertainty quantitatively, and
sometimes even qualitatively.
On the other hand, once a rough estimate of shape of the spectral
function is in hand, $\chi^2$ fit gives more quantitative result
for the parameters of assumed form.
Therefore, complementary use of these methods are preferable for
justification of their applicability and quantitative analysis.
We use MEM for rough estimation of a shape of spectral function,
and $\chi^2$ fit for more quantitative analysis.

In MEM analysis we test several values of $m_0$ for the default model
function, $m(\omega)=m_0 \omega^2$, and fitting ranges
$[t_{min}, t_{max}]$, where $t_{min}=1$ is fixed.
For the $\chi^2$ fit, we adopt two forms for each peak of spectral
function:
$\delta$ function (denoted as {\it pole}) form, and the relativistic
Breit-Wigner ({\it BW}) form,
\begin{equation}
 A(\omega) = \frac{\omega^2 m \gamma R}
                     {(\omega^2 - m^2)^2 + m^2 \gamma^2}.
\end{equation}
Combining them, we apply {\it 2-pole}, {\it 1-BW}, and {\it BW+pole} 
fits to the correlators.
In BW+pole fit the $\delta$ function is used for subtraction of
contribution from larger $\omega$ region (such as an excited state).

\section{Results of analysis}

\begin{figure}[tb]
\vspace{0.3cm} \hspace{0.3cm}
\includegraphics[width=6.6cm]{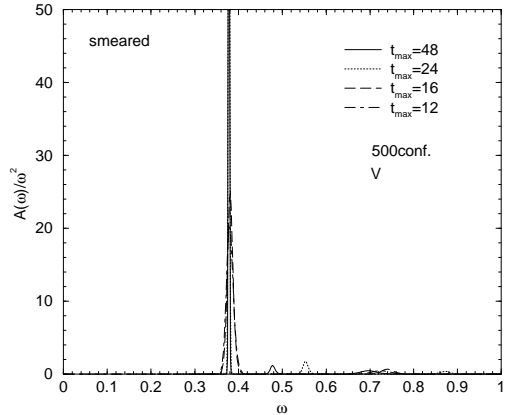}
\vspace{-1.1cm}
\caption{The spectral function in the vector channel determined
by MEM at $T\!=\!0$.}
\label{fig:zero_temp}
\vspace{-0.4cm}
\end{figure}

\begin{figure}[tb]
\hspace*{0.2cm}
\includegraphics[width=6.5cm]{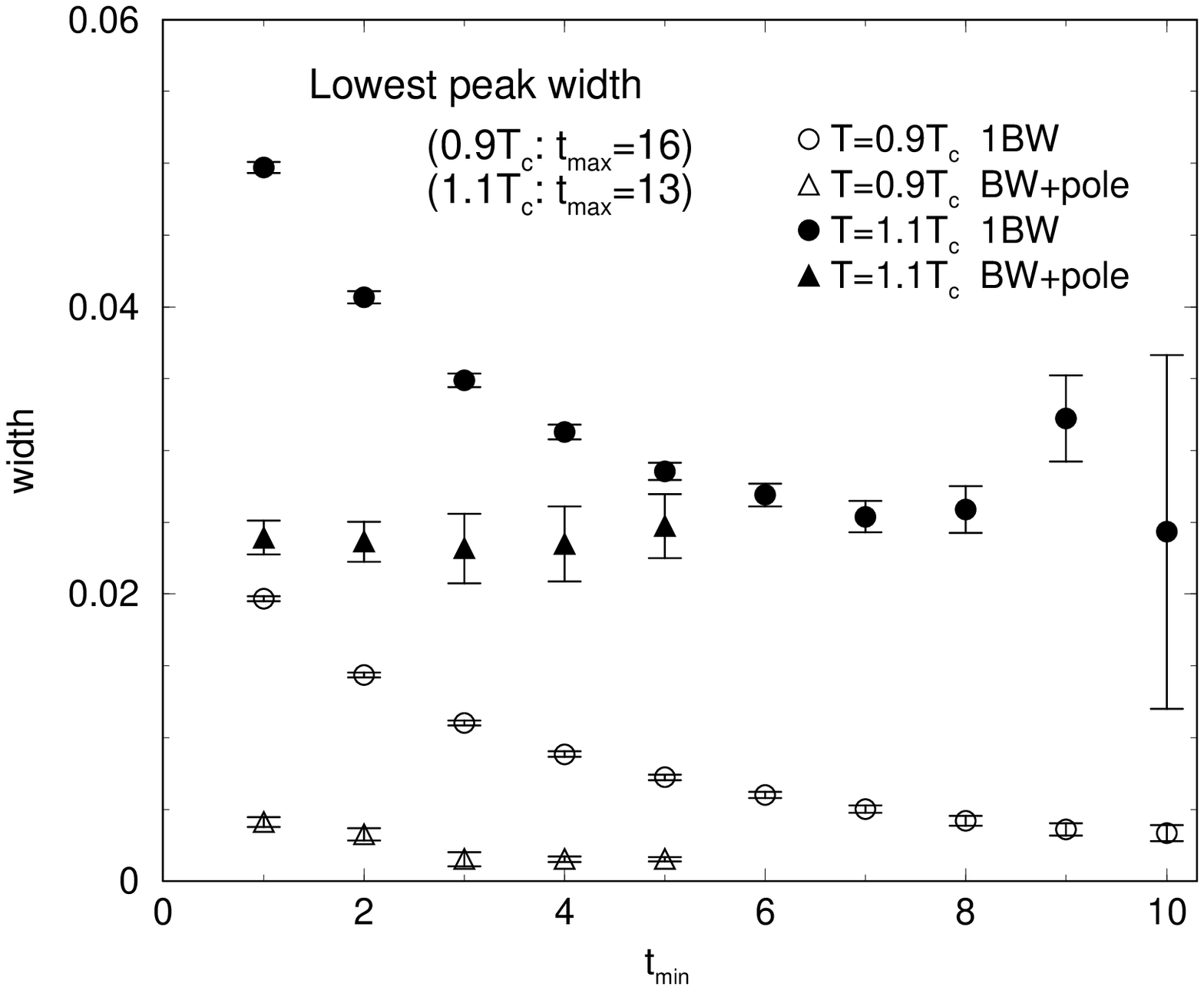}
\vspace{-0.06cm}\\
\hspace*{0.5cm}
\includegraphics[width=6.3cm]{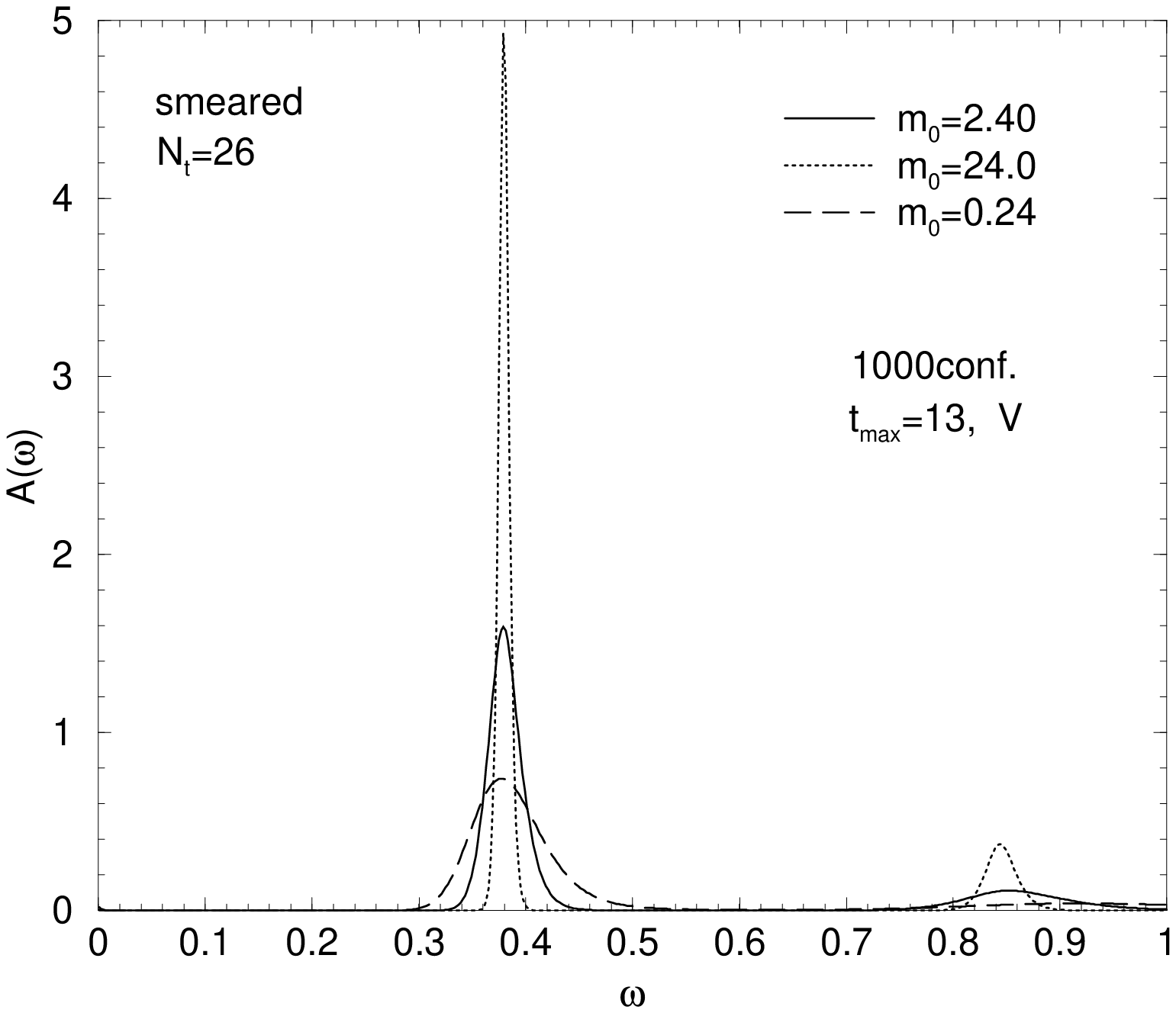}
\vspace{-1.0cm}
\caption{
Result at $T\!>\!0$ for the vector channel.
Top panel shows the result of 1-BW and BW+pole fits
for the ground state peak width at $T \simeq 0.9 T_c$
and $1.1 T_c$.
Bottom panel is the result of MEM at $T \simeq 1.1 T_c$.}
\label{fig:FT}
\vspace{-0.5cm}
\end{figure}

In the following, we show only the result for the vector channel,
while similar result is observed for the pseudoscalar channel.
In order for MEM works at $T\!>\!0$, it should produce at $T\!=\!0$
a structure which is stable under the change of fitting range,
$t_{max}$.
We regard this as a necessary condition for applicability of
MEM to problems at $T\!>\!0$.
Figure~\ref{fig:zero_temp} shows the result of MEM at $T\!=\!0$
with several values of $t_{max}$.
The position of peak corresponding to the ground state is stable,
while the width becomes broader for smaller $t_{max}$.
Although the position of the first excited state changes with $t_{max}$,
we do not consider this seriously, since we are interested only
in the ground state, and the contribution of excited states are small
owing to the smearing of operator.
Therefore, using only restricted numbers of points, which are
inevitable at $T\!>\!0$, MEM can produce a consistent result with the case
of sufficiently large $t_{max}$, at least for a rough estimate of lowest
peak structure.
This is in contrast to the case of a correlator of local operators,
for which our MEM analysis does not produce a stable result under
the same change of $t_{max}$.
The correlator is well fitted to the 2-pole form, and 
both 1-BW and BW+pole fits give the values of width $\gamma$
consistent with zero.

Now we turn to the study of spectral function at $T>0$.
At $0.9 T_c$, MEM gives a similar result as $T\!=\!0$ case.
The BW+pole fit gives almost the same mass as at $T\!=\!0$,
and small value of width (top panel of Fig.~\ref{fig:FT}).
Since this is the most general fit form, we conclude that at
$T\simeq 0.9 T_c$, the ground state has almost the same mass as at
$T\!=\!0$ and almost vanishing width.
The other fits support this conclusion.
The 2-pole fit is rather well applied, while the obtained
mass of the ground state slightly decreases as increasing
$t_{min}$.
The result of 1-BW fit approaches to the consistent value with
BW+pole fit as $t_{min}$ increases.

The result of MEM at $T\simeq 1.1 T_c$ is displayed in Fig.~2 (bottom).
Although the width is larger than $T<T_c$, the result still exhibits
a peak structure around the same energy region as at $T<T_c$.
Therefore we perform the same types of fit analysis as at $T<T_c$.
The 2-pole fit and BW type fits give inconsistent results,
and the latter fits indicate that the spectral function has a peak
with almost the same mass at $T<T_c$ and the width of order of
200 MeV, as shown in Fig.~2 (top).
This result indicate that the charmonium correlator may still have
a nontrivial structure slightly above $T_c$ \cite{Charm}.
In this analysis, we use specific smearing function.
In order to verify that the observed spectral function is not an
artifact \cite{Bielefeld}, comparative study with different smearing
function is necessary, which is now underway.

\end{document}